\documentclass[12pt, fleqn]{article}
\usepackage{graphicx,xcolor,verbatim,ulem}
\usepackage{amssymb,amsfonts,amsmath}
\usepackage{graphics,epsfig}
\usepackage{graphicx}
\usepackage{latexsym,mathtools}
\usepackage{times}
\usepackage{hyperref}
\usepackage{color}
\usepackage{float}
\usepackage{enumerate}
\hypersetup{colorlinks = true, linktocpage = true, bookmarksopen = true, linkcolor = blue, urlcolor=blue, citecolor = blue, allcolors=blue}
\usepackage{authblk}
\usepackage[size=small]{caption}
\usepackage{soul}
\setstcolor{red}

\graphicspath{{./Figures/}}

\usepackage[margin=2.11cm]{geometry}

\newcommand{\be}{\begin{equation}}
\newcommand{\ee}{\end{equation}}
\newcommand{\bc}{\begin{center}}
\newcommand{\ec}{\end{center}}
\newcommand{\bdm}{\begin{displaymath}}
\newcommand{\edm}{\end{displaymath}}
\newcommand{\ds}{\displaystyle}
\newcommand{\p}{\partial}

\newcommand{\bfm}[1]{\mbox{\textbf{#1}}}

\newcommand{\ba}{\begin{array}}
\newcommand{\ea}{\end{array}}
\newcommand{\Dmax}{D_{\text{max}}}
\newcommand{\Dmin}{D_{\text{min}}}

\newcommand{\Davg}{D_{\text{avg}}}
\newcommand{\kmax}{k_{\text{max}}}
\newcommand{\kmin}{k_{\text{min}}}

\newcommand{\kavg}{k_{\text{avg}}}
\newcommand{\rd}{\hat{r}}
\newcommand{\td}{\hat{t}}
\newcommand{\cd}{\hat{c}}

\newcommand{\Pd}{\hat{P}}
\newcommand{\sigmad}{\hat{\sigma}}
\newcommand{\alphad}{\hat{\alpha}}
\newcommand{\Dd}{\hat{D}}
\newcommand{\kd}{\hat{k}}
\newcommand{\Td}{\hat{T}}
\newcommand{\Md}{\hat{M}}

\usepackage[T1]{fontenc}

\begin{document}

\title{\bf Modelling functionalized drug release \\
for a spherical capsule}

\author[1]{ Elliot Carr\footnote{Corresponding author: \href{mailto:elliot.carr@qut.edu.au}{elliot.carr@qut.edu.au}}}
\author[2]{ Giuseppe Pontrelli}
\affil[1]{{\small  School of Mathematical Sciences, Queensland University of Technology (QUT), Brisbane, Australia}}
\affil[2]{{\small Istituto per le Applicazioni del Calcolo - CNR,  Rome, Italy}}

\maketitle
\begin{abstract}
\noindent Advances in material design has led to the rapid development of novel materials with increasing complexity and functions in bioengineering. In particular, functionally graded materials (FGMs) offer important advantages in various fields of application. In this work, we consider a heterogeneous reaction-diffusion model for an FGM spherical drug releasing system that generalizes the multi-layer configuration to arbitrary spatially-variable coefficients. Our model proposes a possible form for the drug diffusivity and reaction rate functions  exhibiting fixed average material properties and a drug release profile that can be continuously varied between the limiting cases of a homogeneous system (constant coefficients) and two-layer system (stepwise coefficients). A hybrid analytical-numerical solution is then used to solve the model, which provides closed-form expressions for the drug concentration and drug release profiles in terms of generalized Fourier series. The resulting concentration and mass profiles show how the release rate can be controlled and continuously varied between a fast (homogeneous) and slow (two-layer) release.
\end{abstract}

\section{Introduction}
Spherical drug carriers are among the most common formulations for a controlled release system. In particular, microcapsules are small spherical particles produced by coating templates constituted of different polymers and using various fabrication strategies \cite{tim}. Although they can be made of a variety of sizes and materials, capsules of interest for most bio-applications have diameters ranging from some nanometers to a few micrometers. Specific examples include liposomes, pellets, nanocontainers, and others \cite{koker}. The effectiveness of polymeric delivery systems can be improved by designing structures with modified material properties that are capable of responding to specific pre-set conditions that prescribe the release of the loaded drug.

One of the approaches generally recognized as effective in the assembly of polymer particles is the layer-by-layer technique. These drug carriers are considered as challenging releasing devices because of their unique multi-layer structural properties \cite{mat}. Another family of drug delivery systems is constituted by stimuli-responsive capsules that control the release of the therapeutic active agents in response to external triggers such as temperature, pH and many others \cite{ton1,maju}. Among other concurrent effects, such as dissolution, polymer swelling and possible degradation, diffusion remains the most important mechanism used to control the release rate from drug delivery systems \cite{ari,gras}. Some experiments, however, show that a fraction of the initial drug loaded is retained within the shell and is never released, due to the specific capacity of the polymer to permanently bind the drug molecules \cite{bar1}. It is common to model this observed phenomenon through first order reaction kinetics \cite{ton1,bar1}. 

Mathematical models of drug release from spherical carriers provide insights 
of mass transport and drug kinetics involved in drug delivery as well as the effect of design parameters, such as the device geometry and drug loading distribution, on the release mechanism  and can significantly reduce the number of experimental studies \cite{kalo}. However, these models depend on so many variables and parameters that, if not appropriately simplified, can raise more questions than useful answers. Analysis of diffusion-controlled system are confined to homogeneous spheres where an exact solution is available \cite{cra}, or to layered capsules \cite{kaoui,carr}, where various mathematical models have been proposed to describe the drug release from this system over the years \cite{tim,ari,kalo,rodr,siep}.

\begin{figure}[t]
\centering
\fbox{\includegraphics[height=0.42\textwidth]{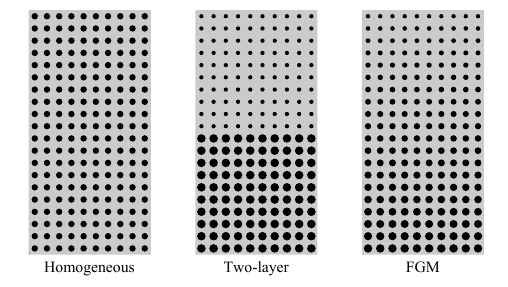}}
\caption{Schematic representation of an FGM with continuous variation of porosity/density compared to an homogeneous material (no variation) and a two-layer material (stepwise variation).}
\label{fig:fig1}
\end{figure}

\begin{figure}[t]
\centering
\fbox{\includegraphics[height=0.4\textwidth]{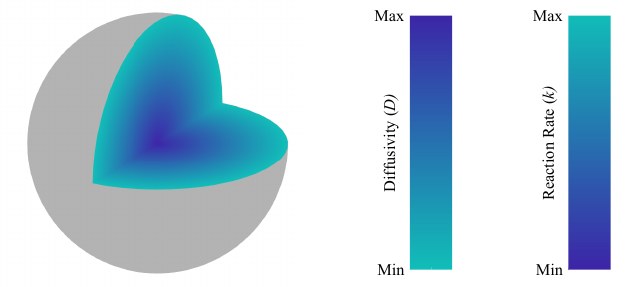}}
\caption{Schematic representation of an FGM spherical capsule \cite{loh}.}
\label{fig:fig2}
\end{figure}

While such configurations are well understood, there is still room for improvement in mechanistic models to control the release mechanism from a drug-loaded sphere. For example, the effect of non-homogeneity represents an important feature that can influence greatly the release properties. Functionally graded materials (FGMs) are a variety of composite materials in which the material properties vary smoothly and continuously (Fig \ref{fig:fig1}). This is in contrast to previous approaches for achieving varying material properties, such as layer-by-layer assembly, where there is an abrupt change in properties from one layer to the next. FGMs, i.e composite materials that have a progressive compositional gradient, are already currently used in a wide range of applications \cite{saleh,shin}. Using today's micro-engineering potential, it is possible to manufacture and control the material properties of the substrate to have the desired smart release properties \cite{koker}. For example, new possibilities are derived from 3D printing technology to manufacture material micro-porosity and density in non-homogeneous substrates \cite{zim,loh}.

A mathematical model of drug release from a thin film FGM has been recently proposed and solved numerically \cite{bre}. In the current work, we propose a reaction-diffusion continuum model to describe drug transport within, and release from, a drug-loaded FGM spherical capsule and develop a hybrid analytical-numerical eigenfunction expansion solution \cite{mik,uni,cot1, cot2, cot4} to handle the spatially-variable coefficients. Our model proposes a possible form for the drug diffusivity and reaction rate functions, which exhibits the same average material properties and can be continuously varied between the limiting cases of a homogeneous system (constant coefficients) and two-layer system (stepwise coefficients)

The rest of the article is organized as follows. In the next section, we present the model equations and boundary conditions that govern the drug mass release from a non-homogeneous FGM spherical system. In section~\ref{sec:analytical_solution}, we present the hybrid analytical-numerical solution methodology leading to a closed-form solution. In section~\ref{sec:results}, drug concentration and drug release profiles are presented for two distinct cases: pure diffusion and reaction-diffusion. Through extensive simulations, we explore the effect of FGM systems on the drug release mechansim by comparing the release profiles to those obtained from standard homogeneous and two-layer systems. Finally, section \ref{sec:conclusions} provides general conclusions and some perspectives for future studies.

\section{Using FGM in releasing spherical particles}
Drug nanocontainers  and  releasing microcapsules are the subject of considerable research effort because of their structural and morphological properties, allowing the synthesis of materials capable of responding to biochemical alterations of the environment \cite{ton1}. Particularly, layer-by-layer polymeric releasing particles have gained increasing interest for their ability to control and tune the release of one or more therapeutic drugs \cite{tsi}. Here, the layers are constituted of different materials having specific physico-chemical characteristics and are customized to allow a selective diffusion and better control the transfer rate \cite{tim}. In the layer-by-layer configuration, a semi-permeable external shell (coating) is often designed to shield and preserve the encapsulated drug from degradation and chemical aggression, and guarantee a more controlled and sustained release \cite{henn}. With the aim of overcoming and generalizing the layered structure, we explore the potential of a material with continuously changing properties. \par

Recently, more attention has been paid to the class of functionally graded materials (FGMs),
 in several fields of application \cite{shin}.
FGMs are a special kind of composite materials in which the microstructural properties 
vary smoothly and continuously 
in space \cite{saleh}. 
In a purely diffusive model, the continuously varying nature of FGMs naturally lends itself to  different
 functional forms of the diffusion coefficient $D(\bfm x)$ in the domain $\Omega$:  
\be
\frac{\partial {c}}{\partial t} = \nabla\cdot\left( D(\bfm x) \nabla c \right) , \qquad   \bfm x \in \Omega, \quad t \in [0, T],\label{heat:eq} 
\ee
where $c(\bfm x,t)$ is the mass volume-averaged concentration of drug. We assume the diffusivity is higher at lower polymer density (inner region) and lower at higher polymer density (outer region), to account for a material that gradually thickens outwards. 

From experiments, however, it is observed that a fraction of the initial loaded drug is retained and never released. A possible explanation of this phenomenon is a chemical reaction due to polymer-drug interaction. In other words, due to long polymeric chains and possible electrostatic interactions, a small percentage of the initial loaded drug is entrapped without being released \cite{ton1}. We model this phenomenon by using reaction kinetics, where the drug molecules travelling through the polymer can potentially be permanently bound with a rate $k(\mathbf{x})$ \cite{bar1}. This generalizes equation (\ref{heat:eq}) to include a first-order reaction term:
\be
\frac{\partial {c}}{\partial t} = \nabla\cdot\left( D(\bfm x) \nabla c \right)  -k(\bfm x) c , \qquad   \bfm x \in \Omega, \quad t \in [0, T],\label{heat:eq1} 
\ee
where $k (\bfm x)$ $[s^{-1}]$ is a space-dependent reaction rate.

\bigskip
\noindent\underline{Drug release from a FGM sphere}\medskip

\noindent We consider a reservoir-type drug carrier with an active agent loaded in a spherical polymeric matrix, which is one of the most common formulations for a controlled release system. The spherical carrier is assumed to have radius $R$ giving domain $\Omega = \{\mathbf{x}\in\mathbb{R}^{3}\,|\,\|\mathbf{x}\|<R\}$ and outer surface $\omega = \{\mathbf{x}\in\mathbb{R}^{3}\,|\,\|\mathbf{x}\|=R\}$, where $\|\cdot\|$ is the Euclidean norm. The interior of the spherical capsule is made of a non-homogeneous FGM, reflecting a customized composition that allows for selective spatially-dependent diffusion and reaction to better control the drug transfer rate (Fig \ref{fig:fig2}). The case of a core-shell capsule is also included in this model, through stepwise diffusivity and reaction functions. As for \textit{in-vitro} experiments, the sphere is immersed in an external ambient medium of a large extent (relative to size of the sphere), taken as semi-infinite.

In the case of an isotropic sphere centered on the origin with a boundary condition on its outer surface, we can assume that net drug diffusion occurs along the radial ($r$) direction only, and thus we restrict our study to a one-dimensional model, as follows:
\begin{align}
&\frac{\partial {c}}{\partial t} =  {1 \over r^2}{\p \over \p r}  \left( r^2 D(r) {\p c  \over \p r} \right)  - k(r) c, \qquad   r \in [0, R], \quad t \in [0, T],\label{eq:dpde}\\ 
&c(r,0) = c_0(r),  \qquad  r \in [0, R],\label{eq:dinit}\\
& {\p c  \over \p r }(0,t) = 0, \qquad  t \in [0, T],   \label{eq:dleftb}\\
&- D(R) {\p c  \over \p r }(R,t) = P \, c(R,t), \qquad  t \in [0, T]. \label{eq:drightb}
\end{align}
The model model permits a space dependent initial concentration (\ref{eq:dinit}) and accounts for a flux resistance (with mass transfer coefficient~$P$) at the external surface (\ref{eq:drightb}) due to the semi-permeable coating~\cite{carr}. 

\bigskip
\noindent\underline{The choice of $D(r)$ and $k(r)$} 
\medskip

\noindent A specific form for $D(r)$ is needed to characterize the material heterogeneity of the FGM. In particular, we assume that the physical medium properties (density or porosity) may change continuously along the radius, being softer in the core and harder towards the surface, leading to a decreasing function $D(r)$ (cf.~Fig \ref{fig:fig2}) that varies from a maximum possible value of $\Dmax$ at the center ($r=0$) to a minimum possible value of $\Dmin$ at the surface ($r=R$). Among a variety of feasible continuous diffusivity functions, we consider
\begin{align}
\label{eq:diff_func}
D(r) = \Dmax + (\Dmin-\Dmax)\left[\frac{1}{2}+\frac{1}{\pi}\arctan\left(\frac{\alpha(r-\sigma)}{R}\right)\right].
\end{align}
This choice is a standard smooth approximation to a two-layer stepwise diffusivity $D(r) = \Dmax + (\Dmin-\Dmax)H(r-\sigma)$ where $H(\cdot)$ is the Heaviside function at $r = \sigma$ with $\alpha>0$ [--] inversely related to the width of the
transition layer and $\sigma$ [cm] denoting the location of the
transition center. 

Our aim is to understand the effect of varying $\alpha$ and $\sigma$ on the drug release profile. To maintain an FGM with the same average diffusive properties, for a specified choice of $\alpha$, we calculate a corresponding value of $\sigma$ so that the average value of $D(r)$ over the spherical capsule is constant:
\begin{align*}
\frac{1}{V(\Omega)}\iiint_{\Omega} D(\mathbf{x})\,\text{d}\mathbf{x} = \Davg,
\end{align*}
which simplifies to
\begin{align}
\label{eq:Davg}
\frac{3}{R^{3}}\int_{0}^{R} r^{2}D(r)\,\text{d}r = \Davg,
\end{align}
in spherical coordinates, when using radial symmetry and $V(\Omega) = 4\pi R^{3}/3$. In our results, we set
\begin{align}
\Davg = \frac{3}{R^{3}}\left[\int_{0}^{R/2} r^{2}\Dmax\,\text{d}r + \int_{R/2}^{R} r^{2}\Dmin\,\text{d}r\right] = \frac{1}{8}\Dmax + \frac{7}{8}\Dmin,
\end{align}
which is the unique value that yields $\sigma \rightarrow R/2$ in the limiting case of a two-layer stepwise medium ($\alpha\rightarrow\infty$). In summary, for a specified choice of $\alpha$, we calculate $\sigma$ by solving the nonlinear equation:
\begin{align}
\label{eq:sigma_eq}
\frac{3}{R^{3}}\int_{0}^{R} r^{2}\left(\Dmax + (\Dmin-\Dmax)\left[\frac{1}{2}+\frac{1}{\pi}\arctan\left(\frac{\alpha(r-\sigma)}{R}\right)\right]\right)\,\text{d}r = \frac{1}{8}\Dmax + \frac{7}{8}\Dmin,
\end{align}
As a consequence of the decreasing diffusivity towards the external surface (due to the thickening material) the drug reaction rate $k(r)$, which is typically proportional to the polymer density/porosity, undergoes a similar radial variation as $D(r)$, but in the opposite direction (cf.~Fig \ref{fig:fig2}), resulting in an increasing function from a minimum possible value of $\kmin$ at the centre ($r=0$) to a maximum possible value of $\kmax$ at the surface ($r=R$) :
\begin{align}
\label{eq:decay_func}
k(r)=\kmin + (\kmax-\kmin)\left[\frac{1}{2}+\frac{1}{\pi}\arctan\left(\frac{\alpha(r-\sigma)}{R}\right)\right],   
\end{align}
where the values of $\alpha$ and $\sigma$ are the same as those used for $D(r)$. As for $D(r)$, the average value of $k(r)$ over the full spherical capsule is constant:
\begin{align*}
\frac{3}{R^{3}}\int_{0}^{R} r^{2}k(r)\,\text{d}r = \kavg=\frac{1}{8}\kmin + \frac{7}{8}\kmax.
\end{align*} 

\section{Solution methodology}
\label{sec:analytical_solution}
\setcounter{equation}{0}
We solve the heterogeneous diffusion model (\ref{eq:dpde})--(\ref{eq:drightb}) by introducing the following non-dimensional variables: 
\begin{align}
\label{eq:dimless_vars1}
&\rd := {r \over R}, \qquad \td := {\Dmax t \over R^2},  \qquad \cd(\rd,\td) := {c(r,t) \over C_0},\\ 
&\cd_{0}(\rd) := \frac{c_0(r)}{C_{0}},\qquad \Td := { \Dmax T \over R^2},\qquad \Pd  := { P  R\over \Dmax},\\
\label{eq:dimless_vars2}
&  \Dd(\rd)  := { D(r)  \over \Dmax}, \qquad  \kd(\rd) := { R^2k(r)   \over \Dmax}, \qquad \alphad := \alpha, \qquad \sigmad := { \sigma \over R},
\end{align}
where $C_{0} = \ds\max_{r\in[0,R]} c_{0}(r)$ and considering the resulting non-dimensional analogue of equations (\ref{eq:dpde})--(\ref{eq:drightb}):
\begin{align}
&\frac{\partial\cd}{\partial\td} =  {1 \over \rd^2}{\p \over \p\rd}  \left( \rd^2 \Dd(\rd) {\p\cd  \over \p\rd} \right)  - \kd(\rd)\cd, \qquad   \rd \in [0, 1], \quad \td \in [0, \Td],\label{eq:pde}\\ 
&\cd(\rd,0) = \cd_{0}(\rd),  \qquad  \rd \in [0, 1],\label{eq:init}\\
& {\p\cd  \over \p \rd }(0,\td) = 0, \qquad  \td \in [0, \Td],   \label{eq:leftb}\\
&- \Dd (1) {\p\cd \over \p\rd}(1,t) = \Pd \, \cd(1,\td), \qquad  \td \in [0, \Td]. \label{eq:rightb}
\end{align}
Equations (\ref{eq:pde})--(\ref{eq:rightb}) constitute a linear problem with spatially-variable coefficients and homogeneous boundary conditions. To solve this problem, we use a hybrid analytical-numerical approach where $\cd(\rd,\td)$ is expanded in terms of eigenfunctions:
\begin{gather}
\label{eq:c_expansion}
\cd(\rd,\td) = \sum_{n=1}^{\infty} T_{n}(\td)X_{n}(\rd).
\end{gather}

\bigskip
\noindent \underline{The space function $X_n(\rd)$}\medskip

\noindent In the solution expansion (\ref{eq:c_expansion}), we let $X_{n}(\rd)$ be the eigenfunctions associated with the following Sturm-Liouville problem:
\begin{align}
\label{eq:Xn_ode}
&\frac{1}{\rd^{2}}\frac{\text{d}}{\text{d}\rd}\left(\rd^{2}\frac{\text{d}X}{\text{d}\rd}\right) = -\lambda^{2} X,\\
\label{eq:Xn_bc}
&- \Dd(1)\frac{\text{d}X}{\text{d}\rd}(1) = \Pd X(1).
\end{align}
\noindent The general solution of equation (\ref{eq:Xn_ode}) is
\begin{align}
\label{eq:12}
X(\rd) = \frac{A\sin(\lambda\rd)}{\rd} + \frac{B\cos(\lambda\rd)}{\rd}.
\end{align}
To ensure this solution remains finite as $r$ tends to zero, we require $B = 0$. Substituting (\ref{eq:12}) into the boundary condition (\ref{eq:Xn_bc}) then yields:
\be
\label{eq:eigen}
A\left[(\Pd-\Dd(1))\sin(\lambda) + \Dd(1)\lambda\cos(\lambda)\right] = 0,
\ee
which has a non-trivial solution ($A\neq 0$) if and only if $\lambda$ is a solution of
\be
\label{eq:eigen1}
(\Pd-\Dd(1))\sin(\lambda) + \Dd(1)\lambda\cos(\lambda) = 0,
\ee
or equivalently:
\be
\label{eq:lambdan}
\tan(\lambda) = \frac{\Dd(1)\lambda}{\Dd(1)-\Pd}.
\ee
\noindent The eigenvalues, denoted by $\lambda_{n}$ for $n\in\mathbb{N}^{+}$, are defined as the positive values of $\lambda$ satisfying (\ref{eq:lambdan}) with the corresponding eigenfunctions given by:
\begin{align}
\label{eq:Xnr}
&X_{n}(\rd) = \frac{2\sqrt{\lambda_{n}}}{\sqrt{2\lambda_{n}-\sin(2\lambda_{n})}}\frac{\sin(\lambda_{n}\rd)}{\rd},
\end{align}
which are orthonormal on the interval $[0,1]$: 
\begin{align}
\label{eq:Xn_orthogonality}
\int_{0}^{1} \rd^{2}X_{n}(\rd)X_{m}(\rd)\,\text{d}\rd = \begin{cases} 0, & \text{if $m\neq n$},\\ 1, & \text{if $m=n$}.\end{cases}
\end{align}

\bigskip
\noindent \underline{The time function $T_n(\td)$}\medskip

\noindent In the solution expansion (\ref{eq:c_expansion}), the time functions $T_n(\td)$ are computed by imposing that (\ref{eq:c_expansion}) satisfy the actual governing equation (\ref{eq:pde}) with space dependent $\Dd(\rd)$ and $\kd(\rd)$. We first substitute (\ref{eq:c_expansion}) into (\ref{eq:pde}) and differentiate to give
\begin{align*}
\sum_{n=1}^{\infty}\frac{\text{d}T_{n}}{\text{d}\td}X_{n}(\rd) = \sum_{n=1}^{\infty}T_{n}(\td)\left[\Dd'(\rd)X_{n}'(\rd) + \frac{\Dd(\rd)}{\rd^{2}}\frac{\text{d}}{\text{d}\rd}\left(\rd^{2}\frac{\text{d}X_{n}}{\text{d}\rd}\right)\right] - \kd(\rd)\sum_{n=1}^{\infty}T_{n}(\td)X_{n}(\rd).
\end{align*}
Next, multiplying both sides of this equation by $\rd^{2}X_{m}(\rd)$ and integrating from $\rd = 0$ to $\rd = 1$, we see that $T_{m}(\td)$ satisfies the following differential equation
\begin{align}
\label{eq:Tm_odes}
\frac{\text{d}T_{m}}{\text{d}\td} &= \sum_{n=1}^{\infty}T_{n}(\td)\left[\int_{0}^{1}\rd^{2}\Dd'(\rd)X_{n}'(\rd)X_{m}(\rd)\,\text{d}\rd - \int_{0}^{1}\rd^{2}(\lambda_{n}^{2}\Dd(\rd) + \kd(\rd))X_{n}(\rd)X_{m}(\rd)\,\text{d}\rd\right], \nonumber \\
&=\sum_{n=1}^{\infty} A_{mn} T_{n}(\td)
\end{align}
after making use of the differential equation (\ref{eq:Xn_ode}) and orthogonality (\ref{eq:Xn_orthogonality}). Equation (\ref{eq:Tm_odes}) identifies
\begin{gather}
\label{eq:Amn}
A_{mn} = \int_{0}^{1} \rd^{2}\Dd'(\rd)X_{n}'(\rd)X_{m}(\rd)\,\text{d}\rd - \int_{0}^{1}\rd^{2}\left[\lambda_{n}^{2}\Dd(\rd) + \kd(\rd)\right]X_{n}(\rd)X_{m}(\rd)\,\text{d}\rd,
\end{gather}
where $X_{n}'(\rd)$, appearing in the definition of $A_{mn}$ (\ref{eq:Amn}), is given by:
\begin{gather*}
X_{n}'(\rd) = \frac{2\sqrt{\lambda_{n}}}{\sqrt{2\lambda_{n}-\sin(2\lambda_{n})}}\left[\frac{\lambda_{n}\cos(\lambda_{n}\rd)}{\rd} - \frac{\sin(\lambda_{n}\rd)}{\rd^2}\right].
\end{gather*}
The appropriate initial condition for the differential equation (\ref{eq:Tm_odes}) is identified by combining the initial condition (\ref{eq:init}) with the expansion (\ref{eq:c_expansion}) 
\begin{align*}
\sum_{n=1}^{\infty}T_{n}(0)X_{n}(\rd) = \cd_{0}(\rd),
\end{align*}
and then applying orthogonality (\ref{eq:Xn_orthogonality}):
\begin{gather}
\label{eq:Tn0}
T_{n}(0) = \int_{0}^{1}\rd^{2}c_{0}(\rd)X_{n}(\rd)\,\text{d}\rd.
\end{gather}
\noindent Assembling the differential equations (\ref{eq:Tm_odes}) and initial condition (\ref{eq:Tn0}) for $m\in\mathbb{N}^{+}$ together yields a matrix system:
\begin{align}
\label{eq:Tn1}
\frac{\text{d}\mathbf{T}}{\text{d}\td} = \mathbf{A}\mathbf{T},\quad \mathbf{T}(0) = \mathbf{T}_{0},
\end{align}
where the entries of $\mathbf{A}$ are defined in equation (\ref{eq:Amn}) and $\mathbf{T}_{0} = [T_{1}(0),T_{2}(0),\hdots]^{T}$. The exact solution of (\ref{eq:Tn1}) is expressed in terms of a matrix exponential
\be
\label{eq:Tn2}
\mathbf{T}(\td) = e^{\td\mathbf{A}}\mathbf{T}_{0},
\ee
with the $n$th entry of $\mathbf{T}(\td)$ defining the time function $T_{n}(\td)$ in the solution expansion (\ref{eq:c_expansion}). 

\bigskip
\noindent \underline{Fraction of drug released}\medskip

\noindent To characterise the release process, we calculate the cumulative fraction of drug released as a function of time, $\hat{M}(t)$. This quantity is obtained by integrating the concentration flux over the outer surface of the spherical capsule and normalizing by the initial mass of drug loaded in the capsule:
\begin{align}
\label{eq:mass_fraction}
\hat{M}(t) = \frac{\ds\int_{0}^{t}\left[\ds\iint_{\omega} \left(-D(\mathbf{x})\nabla c(\mathbf{x},s)\cdot\mathbf{n}\right)\,\text{d}\mathbf{x}\right]\text{d}s}{\ds\iiint_{\Omega}c_{0}(\mathbf{x})\,\text{d}\mathbf{x}},
\end{align}
where $\mathbf{n}$ is the unit vector normal to $\omega$ directed outward from $\Omega$. In spherical coordinates under radial symmetry, $\hat{M}(t)$ simplifies to
\begin{align*}
\hat{M}(t) = \frac{R^{2}\ds\int_{0}^{t}-D(R)\ds\frac{\partial c}{\partial r}(R,s)\,\text{d}s}{\ds\int_{0}^{R}r^{2}c_{0}(r)\,\text{d}r},
\end{align*}
Using the boundary condition at the outer surface (\ref{eq:drightb}) and the dimensionless variables (\ref{eq:dimless_vars1})--(\ref{eq:dimless_vars2})~yields:
\begin{align*}
\hat{M}(\td) = \frac{\ds\int_{0}^{\hat{t}}\hat{P}\hat{c}(1,s)\,\text{d}s}{\ds\int_{0}^{1}r^{2}\hat{c}_{0}(\hat{r})\,\text{d}\hat{r}}.
\end{align*}
Finally, inserting the solution expansion (\ref{eq:c_expansion}) gives the final form for the fraction of drug released
\begin{align}
\label{eq:mass}
\hat{M}(\td) = \frac{\hat{P}\ds\sum_{n=1}^{\infty}U_{n}(\td)X_{n}(\rd)}{\ds\int_{0}^{1}r^{2}\hat{c}_{0}(\hat{r})\,\text{d}\hat{r}},
\end{align}
where the $n$th entry of the vector $\mathbf{U}(\td) = \int_{0}^{\td}\mathbf{T}(s)\,\text{d}s = (e^{\hat{t}\mathbf{A}}-\mathbf{I})\mathbf{A}^{-1}\mathbf{T}_{0}$ defines $U_{n}(\td) = \int_{0}^{\td} T_{n}(s)\,\text{d}s$.

\section{Numerical results}
\label{sec:results}
\setcounter{equation}{0}

We now present numerical results for two distinct cases: \textit{pure diffusion} ($k(r) = 0$) and \textit{reaction-diffusion} ($k(r)>0$). All results are calculated using a uniform initial concentration $c_{0}(r) = C_{0}$ and the parameter values given in Table \ref{tab:phys_param}. We have implemented the analytical solution in MATLAB truncating all infinite series at $N = 150$ terms and using a simple bisection method to solve the eigenvalue equation (\ref{eq:eigen1}). The in-built MATLAB functions \texttt{integral}, \texttt{expm} and \texttt{fzero} are used, respectively, to (i) evaluate the integrals in equations (\ref{eq:Amn}) and (\ref{eq:Tn0}) (ii) compute the matrix exponential $e^{\td\mathbf{A}}$ and (iii) solve the nonlinear equation (\ref{eq:sigma_eq}) for $\sigma$. Further implementation details are available in our code, which can be downloaded from \href{https://github.com/elliotcarr/Carr2023a}{https://github.com/elliotcarr/Carr2023a}.

\begin{table*}[!h]\label{table}
\centering
{
\scalebox{1}
{\begin{tabular}{|p{2cm}|l|r|r|} \hline\hline
 Parameter  &  Description & Value [dim.] & Value [non dim.] \\\hline
 {$R$} & Radius & $ 10^{-4}\,\text{cm}$ & $1$ \\
{$\Dmin$} & Minimum diffusivity &  $10^{-13}\,\text{cm}^2/\text{s}$ & $10^{-2}$ \\
{$\Dmax$} & Maximum diffusivity & $10^{-11}\,\text{cm}^2/\text{s}$ & $1$ \\

{$\Davg$} & Average diffusivity &  $1.3375\cdot 10^{-12}\,\text{cm}^2/\text{s}$ & $0.13375$ \\
{$\kmin$} & Minimum reaction rate &  $8\cdot 10^{-5}\,/\text{s}$ & $0.08$ \\
{$\kmax$ } & Maximum reaction rate & $10^{-4}\,/\text{s}$ & $0.1$ \\
{$\kavg$} & Average reaction rate &  $9.75\cdot 10^{-5}\,/\text{s}$ & $0.0975$ \\
{$P$} & Mass transfer coefficient &  $5 \cdot 10^{-8}\,\text{cm}/\text{s}$ & $0.5$ \\
{$T$} & Maximum time   & $3\cdot 10^{4}\,\text{s}$ & $30$ \\
{$C_0$}& Initial concentration   &   $0.4 \,\text{mol}/\text{cm}^3$ & $1$ \\
 \hline\hline
\end{tabular} }}  
\caption{Parameters of the problem.}\label{tab:phys_param}
\end{table*}

\begin{figure}[!t]
\fbox{\includegraphics[width=0.98\textwidth]{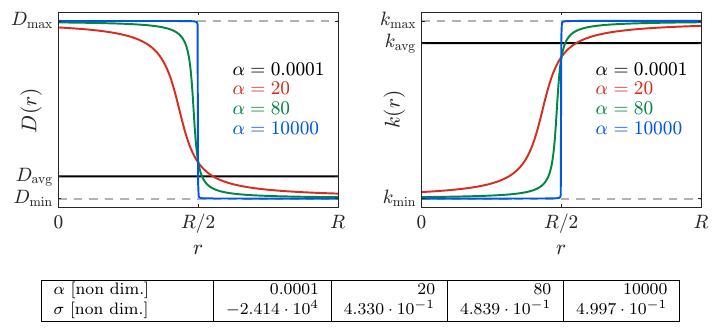}}
\caption{(left) Diffusivity functions $D(r)$ (\ref{eq:diff_func}) used for both the \textit{pure diffusion} and \textit{reaction-diffusion} cases. (right) Reaction rate functions $k(r)$ (\ref{eq:decay_func}) used for the \textit{reaction-diffusion} case. For each value of $\alpha$, the table gives the corresponding value of $\sigma$ satisfying equation (\ref{eq:sigma_eq}).}
\label{fig:diff_funcs}
\end{figure}

Figure \ref{fig:diff_funcs} displays the diffusivity function $D(r)$ and reaction rate function $k(r)$ for the parameter values in Table \ref{tab:phys_param} and four different choices of $\alpha$ and $\sigma$. Observe that the smallest value of $\alpha$ accurately captures the case of a homogeneous medium (e.g. $D(r) = \Davg$), while the largest value of $\alpha$ accurately captures the case of a two-layer medium (e.g. $D(r) = \Dmax$ if $0 < r < R/2$ and $D(r) = \Dmin$ if $R/2 < r < R$).

\begin{figure}[p]
\def\figw{0.34\textwidth}
\centering
\includegraphics[width=1.0\textwidth]{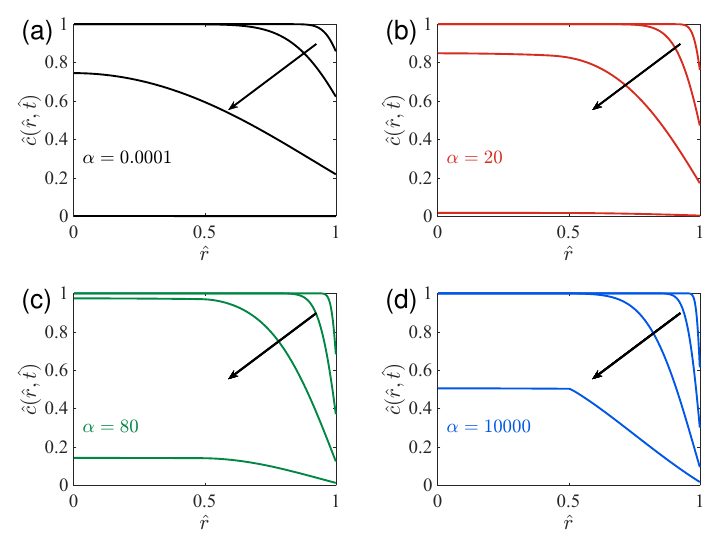}
\caption{Drug concentration as a function of radius (\ref{eq:c_expansion}) for the \textit{pure diffusion} case. Profiles are shown at four distinct times, $\td = 10^{-2},10^{-1},10^{0},10^{1}$, with the black arrow indicating the direction of increasing time.}
\label{fig:conc_pure_diffusion}

\vspace{1.2cm}
\def\figw{0.34\textwidth}
\centering
\includegraphics[height=\figw]{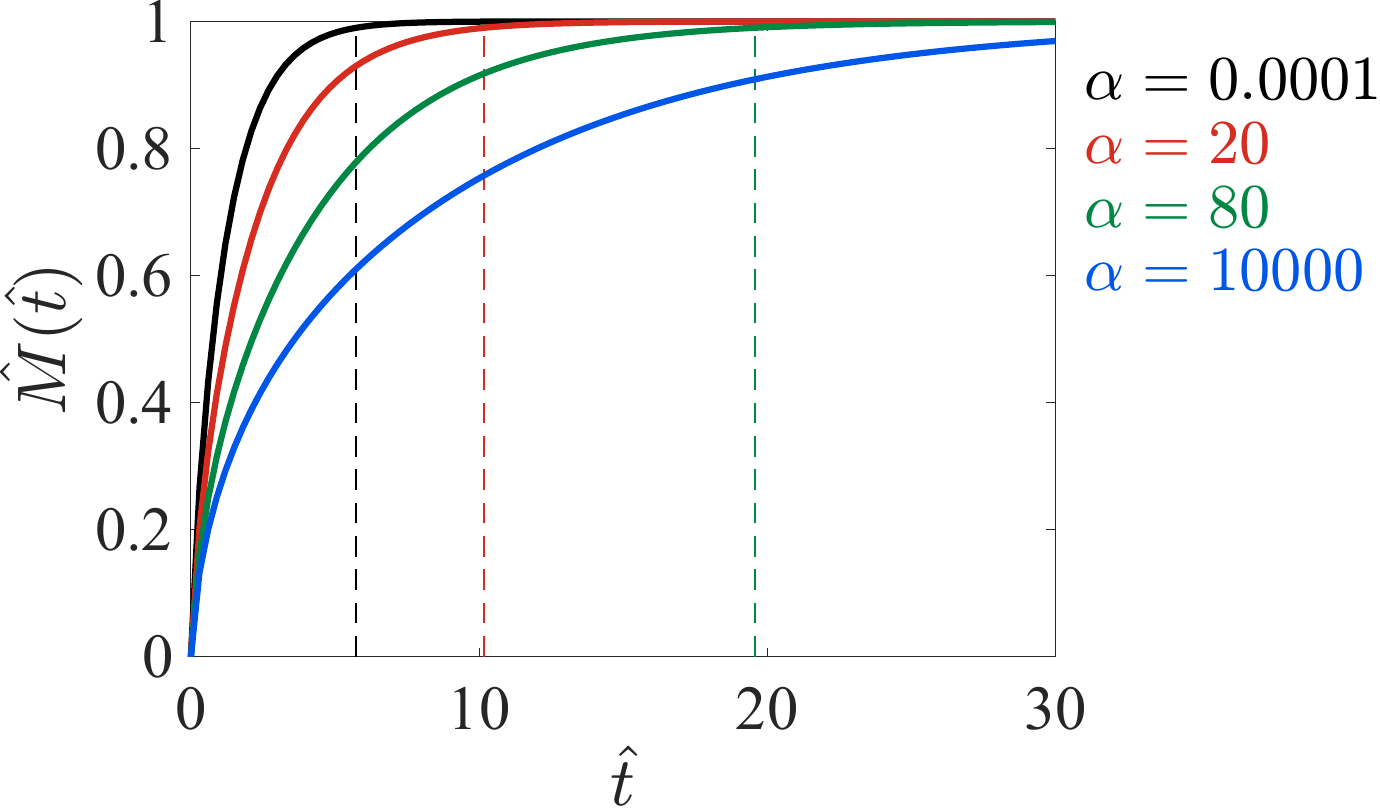}\\
\caption{Cumulative fraction of drug released as a function of time (\ref{eq:mass}) for the \textit{pure diffusion} case. Vertical dashed lines indicate approximate release times corresponding to the time when 99\% of the total released mass has been released, i.e., the time when $\smash{\Md(\td) = 0.99\lim\limits_{\td\rightarrow\infty}\Md(\td)}$. The release time for $\alpha = 10000$ exceeds $30$ and is not shown.}
\label{fig:mass_pure_diffusion}
\end{figure}

For the case of pure diffusion, drug is released differently depending on the parameter $\alpha$ (Figs \ref{fig:conc_pure_diffusion}, \ref{fig:conc_field1} and \ref{fig:conc_field2}) and these results demonstrate the wide variety of concentration profiles using FGMs. Actually, in some circumstances, maintaining local drug concentrations within some defined therapeutic range is desirable, while in  other cases  a {\it burst} of drug is required. If the release is not controlled appropriately, this can lead to periods where toxic and/or sub-therapeutic concentrations are achieved. The inclusion of the spatially-varying diffusivity (\ref{eq:diff_func}) provides greater control over the drug release profile (through the single tuning parameter $\alpha$) while maintaining the same average material properties (\ref{eq:Davg}). In Fig \ref{fig:mass_pure_diffusion}, we see that the fraction of drug released increases more rapidly when decreasing the value of $\alpha$, giving rise to a quicker overall drug delivery for small values of $\alpha$ and a more-sustained release for large values of $\alpha$. The {\it initial burst} of dose at small $\alpha$ may be beneficial when a rapid delivery, rather than a delayed sustained release, is desired. 

\begin{figure}[p]
\def\figw{0.34\textwidth}
\centering
\includegraphics[width=1.0\textwidth]{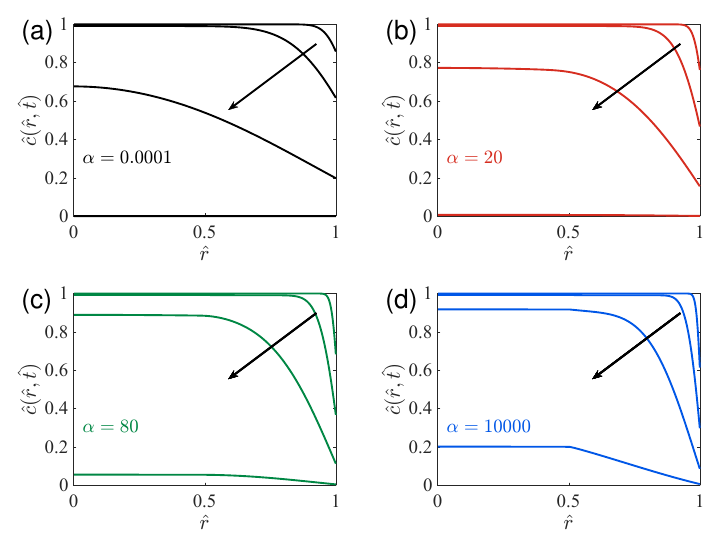}
\caption{Drug concentration as a function of radius (\ref{eq:c_expansion}) for the \textit{reaction-diffusion} case. Profiles are shown at four distinct times, $\td = 10^{-2},10^{-1},10^{0},10^{1}$, with the black arrow indicating the direction of increasing~time.}
\label{fig:conc_reaction_diffusion}

\vspace{1.2cm}
\def\figw{0.34\textwidth}
\centering
\includegraphics[height=\figw]{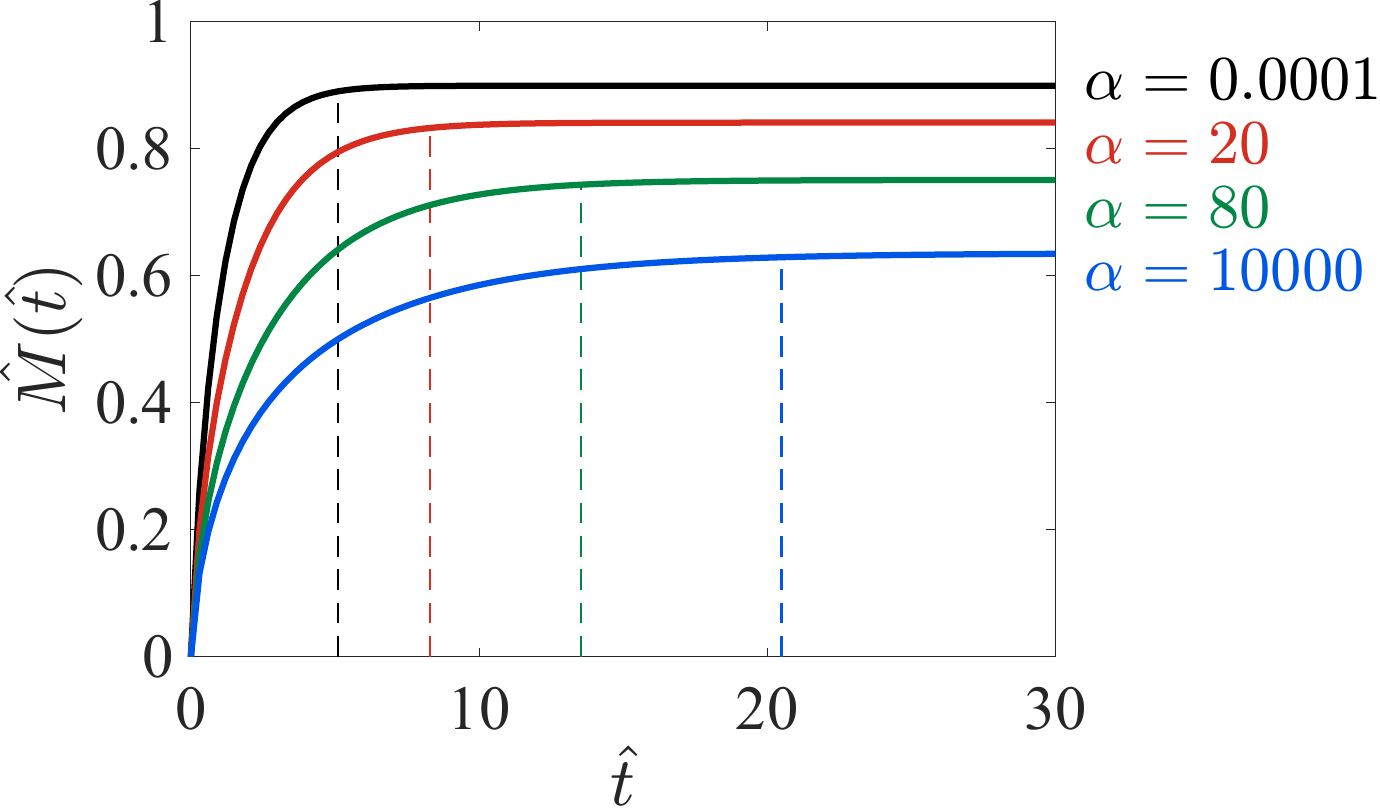}\\
\caption{Cumulative fraction of drug released as a function of time (\ref{eq:mass}) for the \textit{reaction-diffusion} case. Vertical dashed lines indicate approximate release times corresponding to the time when 99\% of the total released mass has been released, i.e., the time when $\Md(\td) = 0.99\lim\limits_{\td\rightarrow\infty}\Md(\td)$.}
\label{fig:mass_reaction_diffusion}
\end{figure}

\begin{figure}[p]
\centering
\def\scal{0.4}
\fbox{\includegraphics[width=0.98\textwidth]{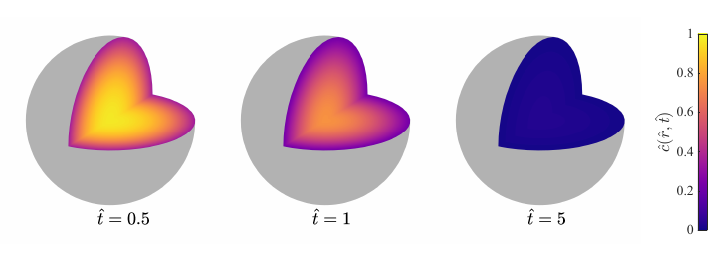}}
\caption{Drug concentration distribution (\ref{eq:c_expansion}) over time for the \textit{pure diffusion} case ($\alpha = 0.0001$).}
\label{fig:conc_field1}
\end{figure}

\begin{figure}[p]
\centering
\def\scal{0.4}
\fbox{\includegraphics[width=0.98\textwidth]{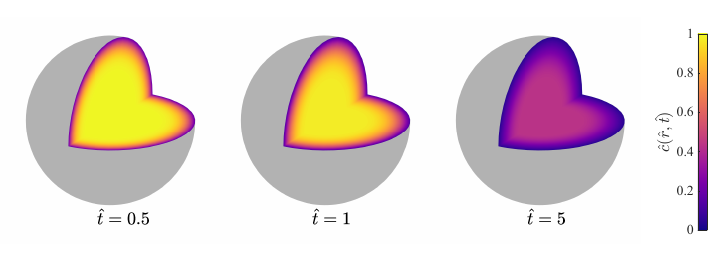}}
\caption{Drug concentration distribution (\ref{eq:c_expansion}) over time for the \textit{pure diffusion} case ($\alpha = 80$).}
\label{fig:conc_field2}
\end{figure}

\begin{figure}[p]
\centering
\def\scal{0.4}
\fbox{\includegraphics[width=0.98\textwidth]{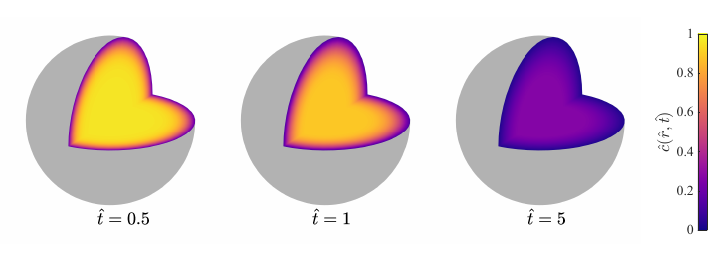}}
\caption{Drug concentration distribution (\ref{eq:c_expansion}) over time for the \textit{reaction diffusion} case ($\alpha = 80$).}
\label{fig:conc_field3}
\end{figure}

Let us now consider the case of reaction diffusion. When reaction is included in the model, the capsule retains a fraction of mass that remains bound to the polymer and is never released. This yields concentration profiles (Figs \ref{fig:conc_reaction_diffusion} and \ref{fig:conc_field3}) that decrease more rapidly than the corresponding profiles obtained for pure diffusion (Figs \ref{fig:conc_pure_diffusion} and \ref{fig:conc_field2}). Since the total cumulative released mass is less than the initial mass of drug loaded in the capsule (cf. equation (\ref{eq:mass_fraction})), the cumulative fraction of drug released asymptotes to a value less than one (Fig \ref{fig:mass_reaction_diffusion}). Comparing the drug release profiles for reaction-diffusion (Fig \ref{fig:mass_reaction_diffusion}) with the corresponding profiles for pure diffusion (Fig \ref{fig:mass_pure_diffusion}) it is clear that (i) the presence of reaction results in decreased release times and (ii) the total released mass decreases for increasing values of $\alpha$ (Fig \ref{fig:mass_reaction_diffusion}). Hence, the inclusion of reaction in the model provides further control of the drug release profile through the tuning parameter $\alpha$. In summary, our model suggests that different functional forms for the diffusivity and reaction rate are able to provide a variety of drug release characteristics different from those provided by the limiting cases of a homogeneous system (constant coefficients) and two-layer system (stepwise coefficients).

\section{Conclusions}
\label{sec:conclusions}
Polymeric engineered materials have been exploited in a range of different application in biomedicine, aerospace and material science and can be useful in the pharmaceutical industry in the field of drug delivery. With recent advances in bioengineering, novel functionally graded materials (FGMs) have been introduced for the development of drug releasing devices and systems. They contribute to the tailoring of material for optimal drug administration including targeted release and customizability. 

The goal of the present study was to elucidate the potential transport mechanism  and the drug kinetics behavior due to the diffusion and reaction shape-material functions, providing guidance for design micro-structure of polymer platforms and capsules development. A variety of space-dependent reaction-diffusion FGM shapes has been proposed and implemented. The hybrid analytical-numerical solution improves the understanding of the mass transfer from an FGM capsule, including in the presence of a binding reaction.

However, it is important to recognize some limitations of the present one-dimensional model. Drug dynamics in the release medium outside the capsule are ignored, so that the interactions between the capsule and the medium are represented entirely by the boundary condition at the external surface. Diffusion coefficients are also assumed to be independent of concentration. For most practical applications such assumptions are reasonable, and therefore, the model results may be helpful in the evaluation of drug kinetics and may provide new pathways for smarter delivery systems. The strategy proposed here, once calibrated, can be utilized in a predictive way to limit the number of experiments. Thus, by showing the correlation between properties of the drug kinetics and material function variables, our model can be used to determine and optimize the processing parameters to ensure a controlled drug delivery within a certain time. 

\section*{Acknowledgments}
G.P. thanks F. De Monte, R. Cotta, C.P. Naveira-Cotta for many useful discussions.  This research has been partly funded by Italian MIUR project `3D-Phys' (Grant No. PRIN 2017PHRM8X).

\section*{Data Availability}
MATLAB code implementing the analytical solution and reproducing the results of the paper is available on GitHub: \href{https://github.com/elliotcarr/Carr2023a}{https://github.com/elliotcarr/Carr2023a}.

\section*{CRediT authorship contribution statement}
\textbf{Elliot J. Carr:} Methodology, Software, Validation, Formal analysis, Investigation, Data curation, Writing - original draft, Writing - review \& editing, Visualization. \textbf{Giuseppe Pontrelli:} Conceptualization, Methodology, Investigation, Writing - original draft, Writing - review \& editing.

\end{document}